\documentclass[twocolumn]{aastex62}

\usepackage{soul}
\usepackage{amsfonts,amsbsy}
\usepackage{mathrsfs}
\usepackage{bm} 

\usepackage{graphicx} 
\usepackage{dcolumn} 

\usepackage{aas_macros}

\usepackage{xcolor}
\usepackage{color}

\usepackage{url}
\usepackage{multirow}

\usepackage{ulem}
\usepackage{enumerate}
\usepackage{textcmds}
\usepackage{mathtools}
\usepackage{booktabs}
\usepackage{tabularx}
\usepackage{lineno}
\usepackage{subfigure}
\usepackage{graphicx}

\newcolumntype{p}{D{,}{\pm}{-1}}

\renewcommand{\figurename}{Fig.}
\renewcommand{\tablename}{Tab.}

\renewcommand{\arraystretch}{1.5}

\begin{document}
	
	\title[]{Prospects for detection rate of very-high-energy $\gamma$-ray emissions from short $\gamma$-ray bursts with the HADAR experiment}
	
	
	\correspondingauthor{Ming-Ming Kang, Yi-Qing Guo, Tian-Lu Chen}
	\email{kangmm@scu.edu.cn, guoyq@ihep.ac.cn, chentl@ihep.ac.cn}

	\author{Qi-Ling Chen}

	\affiliation{College of Physics, Sichuan University, Chengdu 610064, China}
	\author{Pei-Jin Hu}

	\affiliation{College of Physics, Sichuan University, Chengdu 610064, China}
	\author{Jing-Jing Su}
	\affiliation{College of Physics, Sichuan University, Chengdu 610064, China}
	
	\author{Ming-Ming Kang}

	\affiliation{College of Physics, Sichuan University, Chengdu 610064, China}
	
	\author{Yi-Qing Guo}
	\affiliation{Key Laboratory of Particle Astrophysics, Institute of High Energy Physics, Chinese Academy of Sciences, Beijing 100049, China}\affiliation{University of Chinese Academy of Sciences, 19 A Yuquan Rd, Shijingshan District, Beijing 100049, China}

	\author{Tian-Lu Chen}
	\affiliation{The Key Laboratory of Cosmic Rays (Tibet University), Ministry of Education, Lhasa 850000, China}

	\author{Dan-Zeng Luo-Bu}
	\affiliation{The Key Laboratory of Cosmic Rays (Tibet University), Ministry of Education, Lhasa 850000, China}
	\author{Yu-fan Fan}
	\affiliation{The Key Laboratory of Cosmic Rays (Tibet University), Ministry of Education, Lhasa 850000, China}
	\author{You-Liang Feng}
	\affiliation{The Key Laboratory of Cosmic Rays (Tibet University), Ministry of Education, Lhasa 850000, China}
	
	\author{Qi Gao}
	\affiliation{The Key Laboratory of Cosmic Rays (Tibet University), Ministry of Education, Lhasa 850000, China}
	\author{Quan-Bu Gou}
	\affiliation{Key Laboratory of Particle Astrophysics, Institute of High Energy Physics, Chinese Academy of Sciences, Beijing 100049, China}

	\author{Hong-Bo Hu}
	\affiliation{Key Laboratory of Particle Astrophysics, Institute of High Energy Physics, Chinese Academy of Sciences, Beijing 100049, China}
	\affiliation{University of Chinese Academy of Sciences, 19 A Yuquan Rd, Shijingshan District, Beijing 100049, China}

	\author{Hai-Jin Li}
	\affiliation{The Key Laboratory of Cosmic Rays (Tibet University), Ministry of Education, Lhasa 850000, China}
	\author{Cheng Liu}
	\affiliation{Key Laboratory of Particle Astrophysics, Institute of High Energy Physics, Chinese Academy of Sciences, Beijing 100049, China}
	\author{Mao-Yuan Liu}
	\affiliation{The Key Laboratory of Cosmic Rays (Tibet University), Ministry of Education, Lhasa 850000, China}
	\author{Wei Liu}
	\affiliation{Key Laboratory of Particle Astrophysics, Institute of High Energy Physics, Chinese Academy of Sciences, Beijing 100049, China}

	\author{Xiang-Li Qian}
	\affiliation{School of Intelligent Engineering, Shandong Management University, Jinan 250357, China}
	\affiliation{The Key Laboratory of Cosmic Rays (Tibet University), Ministry of Education, Lhasa 850000, China}
	\author{Bing-Qiang Qiao}
	\affiliation{Key Laboratory of Particle Astrophysics, Institute of High Energy Physics, Chinese Academy of Sciences, Beijing 100049, China}
	\author{Hui-Ying Sun}
	\affiliation{School of Intelligent Engineering, Shandong Management University, Jinan 250357, China}
	\author{Xu Wang}
	\affiliation{School of Intelligent Engineering, Shandong Management University, Jinan 250357, China}
	\author{Zhen Wang}
	\affiliation{Tsung-Dao Lee Institute, Shanghai Jiao Tong University, Shanghai 200240, China}
	\author{Guang-Guang Xin}
	\affiliation{School of Physics and Technology, Wuhan University, Wuhan 430072, China}
	\author{Yu-Hua Yao}
	\affiliation{College of Physics, Chongqing University, Chongqing 401331, China}
	\author{Qiang Yuan}
	\affiliation{Key Laboratory of Dark Matter and Space Astronomy, Purple Mountain Observatory, Chinese Academy of Sciences, Nanjing 210008, China}
	\author{Yi Zhang}
	\affiliation{Key Laboratory of Dark Matter and Space Astronomy, Purple Mountain Observatory, Chinese Academy of Sciences, Nanjing 210008, China}
	
	\author{Bing Zhao}
	\affiliation{Key Laboratory of Particle Astrophysics, Institute of High Energy Physics, Chinese Academy of Sciences, Beijing 100049, China}
	
	\begin{abstract}
	The observation of short gamma ray bursts (SGRBs) in the TeV energy range plays an important role in understanding the radiation mechanism and probing new areas of physics such as Lorentz invariance violation. However, no SGRB has been observed in this energy range due to the short duration of SGRBs and the weakness of current experiments. New experiments with new technology are required to detect sub-TeV SGRBs. In this work, we  observe the very high energy (VHE) $\gamma$-ray emissions from SGRBs and calculate the annual detection rate with the High Altitude Detection of Astronomical Radiation HADAR (HADAR) experiment. First, a set of pseudo-SGRB samples is generated and checked using the observations of Fermi-GBM, Fermi-LAT, and SWIFT measurements. The annual detection rate is calculated from these SGRB samples based on the performance of the HADAR instrument. As a result, the HADAR experiment can detect 0.5 SGRB per year if the spectral break-off of $\gamma$-rays caused by the internal absorption is larger than 100 GeV. For a GRB09010-like GRB in HADAR's view, it should be possible to detect approximately 2000 photons considering the internal absorption. With a time delay assumption due to the Lorentz invariance violation effects, a simulated light curve of GRB090510 has evident energy dependence. We hope that the HADAR experiment can perform the SGRB observations and test our calculations in the future.
	\end{abstract}
	
	\section{Introduction}
	\label{sec:intro}
	\par
	GRBs are some of the most powerful explosions in the universe and can last from 10 ms to several hours as prompt emissions, releasing most of  the energy in the form of photons from 30 keV to a few MeV. The properties of GRBs mainly include two components of temporal and spectral information. First, from the point view of the temporal property, its prompt phase has a bimodal duration distribution separated at $t\sim 2$ s, indicating that there are two different groups as the long GRBs (LGRBs) with $t_{LGRB} > 2\ s$ and SGRBs with $t_{SGRB} <2\ s$ \citep{kouveliotou1993identification,meszaros2006gamma}. LGRBs are likely formed by the collapses of massive star cores \citep{woosley1993gamma,macfadyen1999collapsars}, while SGRBs are thought to come from the coalescence of compact binary systems, such as binary neutron stars (BNS) and neutron star-black hole (NS-BH) systems \citep{blinnikov1984exploding,paczynski1986gamma,eichler1989nucleosynthesis}. The associated events GW170817/GRB170817A \citep{abbott2017gravitational} from a BNS merger make the grounded assumption that BNS mergers are sources (or parts of the sources) of SGRBs.  For a spectrum, a two-component structure has been revealed above tens of MeV. At low energy, an experiential Band-function can describe observations well \citep{band1993batse,band2003comparison}, which is possibly induced by the synchrotron radiation process \citep{oganesyan2017detection,oganesyan2018characterization,oganesyan2019prompt,ravasio2018consistency}. At high energy, the model predicts that another component generated from the inverse-Compton (IC) process should exist \citep{meszaros1994delayed,dermer2000beaming,sari2001synchrotron,zhang2001high}. In the past few decades, great achievements have been made in experimental measurements to study temporal and spectral characteristics. It is evident that the measurements are relatively rich at low energy, but observations in the TeV band are more desirable. 
 
	The milestone progress was achieved in observations of $\gamma$-ray emissions in the TeV band in recent years. The MAGIC (Major Atmospheric Gamma Imaging Cherenkov) and H.E.S.S. (High Energy Stereoscopic System) experiments discovered sub-TeV $\gamma$-ray emissions in GRB190114C and GRB180720B \citep{veres2019observation,abdalla2019very}.  Following these observations, the HESS experiments detected photons with energies of up to 4 TeV from GRB190829A, between 4 and 56 hr after the trigger time \citep{hess2021revealing}. The brightest GRB221009A was observed in a multi-band just recently. It is very exciting that the highest photons can reach 18 TeV \citep{huang2022lhaaso}. This new discovery opened the observation window of $\gamma$-ray emissions of GRBs in the TeV band and confirmed the new radiation mechanism expected by the model, named the IC process \citep{meszaros1994delayed,dermer2000beaming,sari2001synchrotron,zhang2001high}. Those  discoveries 
	shined a new light on the central engine and radiation mechanism and probed new physics such as LIV and Dark Matter (DM) \citep{amelino1998tests,nakagawa2023axion,gonzalez2023grb}. It is unfortunate that the $\gamma$-ray emissions at TeV energy were only observed at the afterglow phase of LGRBs. No prompt emission has yet been observed for GRBs at this energy band, let alone SGRBs. Despite this, 
	TeV $\gamma$-ray emissions from SGRBs can play very important roles in understanding LIV because of the short duration of SGRBs. Therefore, scientists are looking forward to the $\gamma$-ray observation results in the TeV energy region.
	
	Space-borne and ground-based experiments have both contributed to the study of LIV. GRB090510 is a typical bright SGRB among high-energy observations, with a photon observed at 31 GeV arriving at 0.829 s after the trigger of Fermi-GBM. This SGRB provides conditions to constrain the LIV effect and provides a lower limit for the linear modification of the photon dispersion relationship, which is $E_{QG,1} > 1.49\times10^{19} \ \text{GeV}$\citep{abdo2009limit},  which is the best down-limit . However, only one photon is detected above 10 GeV by Fermi-LAT in this work\citep{abdo2009limit}, and the event number with high statistics is better for confirmation of this important result. The shortage of space-borne experiments includes a limited effective area of $\sim$1 $m^2$, which leads to insufficient sensitivity in the high-energy range. Imaging atmospheric Cherenkov telescopes (IACT) experiments have high sensitivity above 100 GeV energy, but the fatal weakness for SGRBs is their field of view (FOV) with a value of only several degrees \citep{bolmont2014camera, aleksic2016major, cta2011design}. However, the Cherenkov Telescope Array (CTA) will detect promt emissions of SGRBs when they are slewwd to the source before the BNS mergers \citep{banerjee2022detecting}. The traditional arrays of CRs, such as HAWC (High-Altitude Water Cherenkov) and LHAASO (Large High Altitude Air Shower Observatory), have good performance above TeV energy and wide FOV, but these arrays cannot be compared with IACT detectors at the 100 GeV energy range. 
	The optimal choice is an IACT with a wide FOV such as the HADAR project. Therefore, it is necessary to know the detection capability of SGRBs for the HADAR experiment. In this work, the annual detection rate based on a set of pseudo-SGRB samples is generated and checked with the observations of space-borne experiments.
 
	Our work in this paper is organized as follows. First, we provide a brief introduction to the HADAR experiment and its performance parameters in section 2. Then, to predict the detection rate of SGRBs with HADAR as reliably as possible, we model the pseudo-SGRBs and the simulations of the detecting process with consideration of the performance of detectors for both Fermi-GBM (for consistency check) and HADAR. The simulation methods and results are shown in Sections 3 and 4, respectively. In Section 4, we also describe the simulation of the observational results of GRB090510 in HADAR's case, given as a light curve with consideration of LIV effects. In Section 5, we describe the conclusion of our findings.

	\section{HADAR Experiment}
	\label{sec:HADAR}
	
	\begin{figure*}[!htb]
		\centering
		\includegraphics[width=0.45\textwidth, height=0.3\textheight]{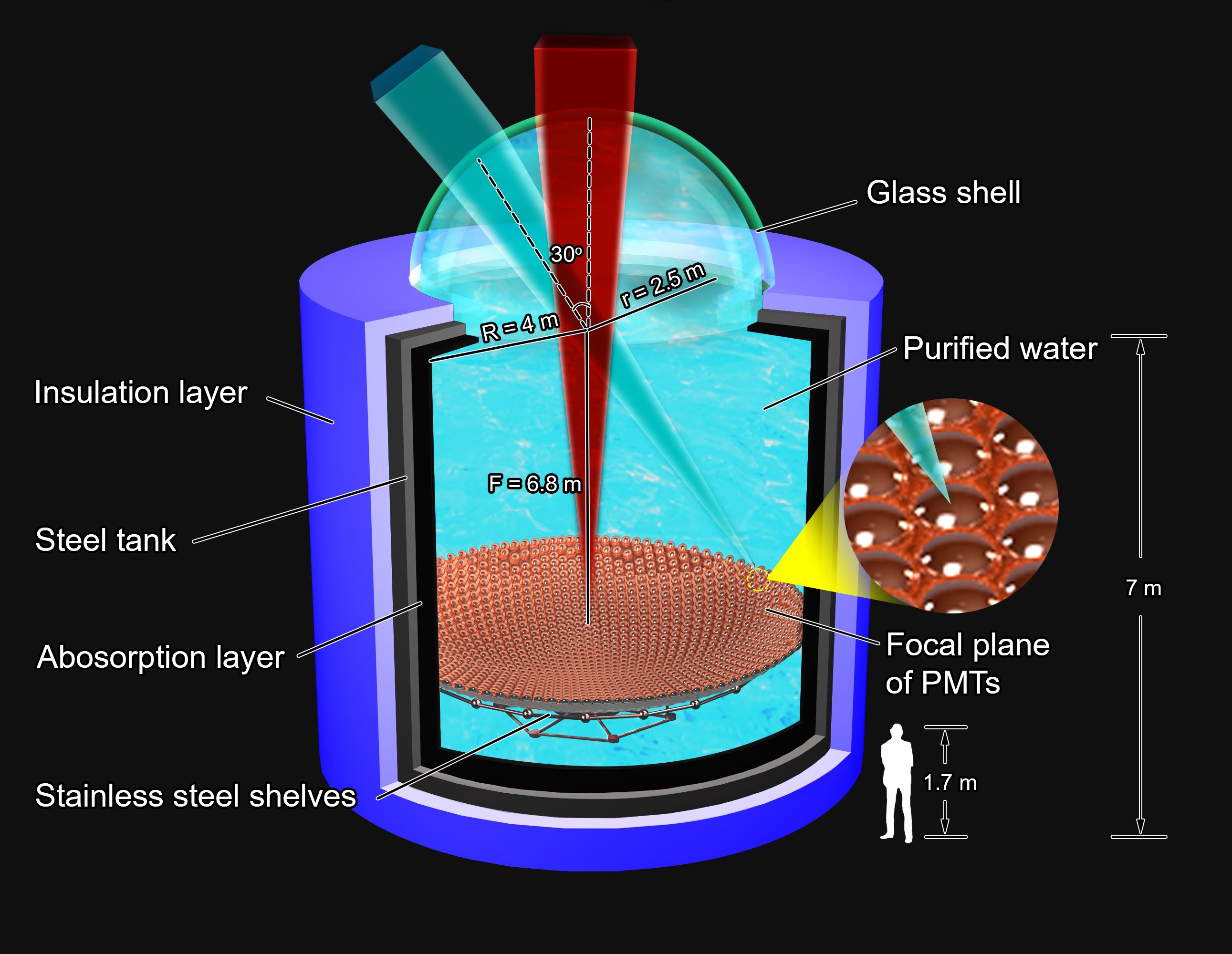}
		\includegraphics[width=0.45\textwidth, height=0.3\textheight]{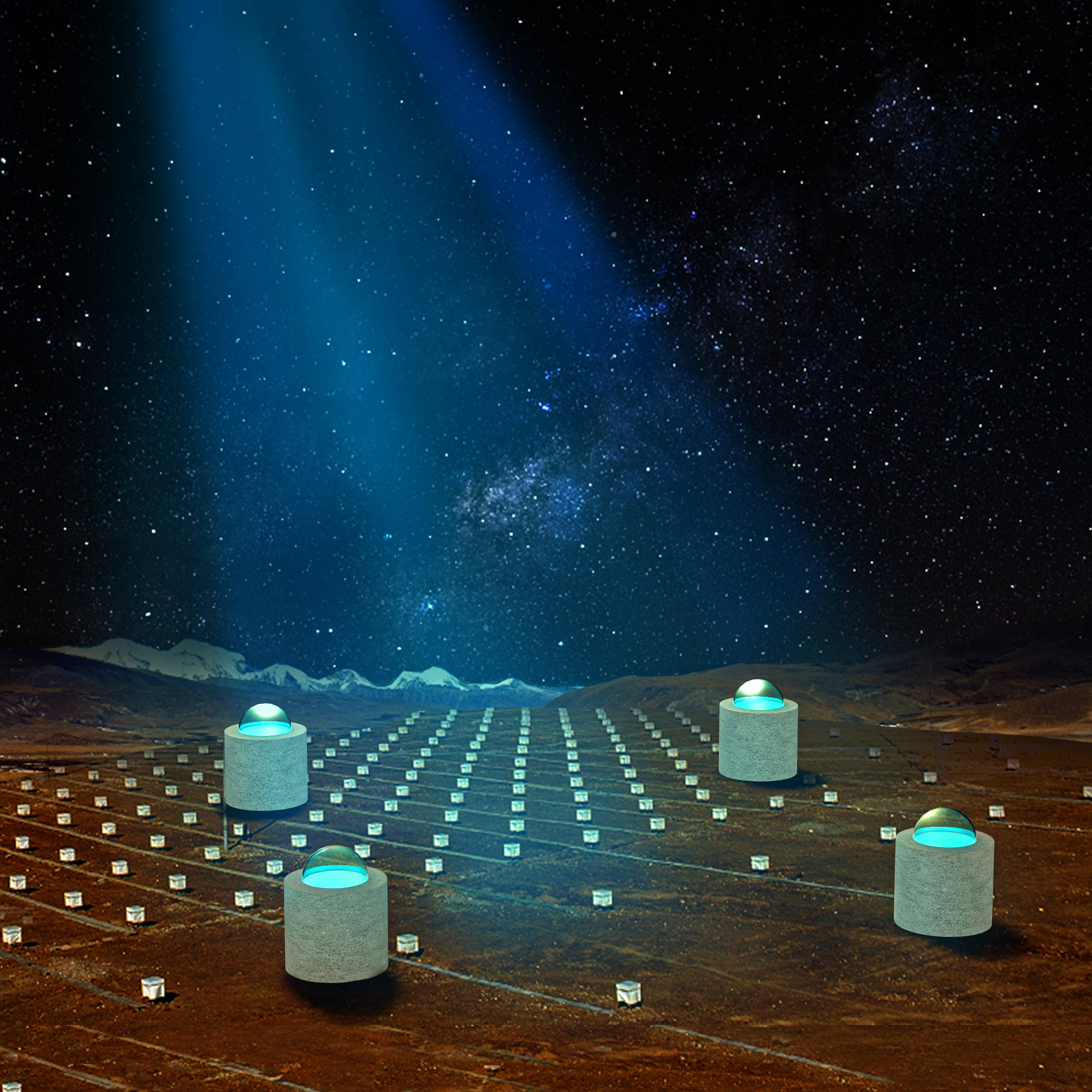}
		\caption{\label{fig:experiment} The HADAR project. Left: Layout of the HADAR array with a water lens. Right: Profile design of each telescope.}
	\end{figure*}
	
	The HADAR project is proposed for construction in YangBaJing, Tibet of China at the altitude of $4300 \ \text{m}$. 
	As shown in Figure \ref{fig:experiment}, this project has a hybrid array, which includes the two parts of the water lens and traditional scintillators. 
	
	The water lens array is composed of four of the same telescopes with a diameter of 5 m and a distance of 100 m between two telescopes. The right panel of the figure \ref{fig:experiment} presents the detailed profile structure of a single water lens. The profile structure mainly consists of an acrylic spherical cap lens with a diameter of 5 m, a cylindrical tank with a radius of 4\ m and a height of 7\ m, and a camera with an array of 18,961 Photomultiplier tubes (PMTs) at each 5 cm in diameter. The steel tank, which contains an absorption layer in the inside wall and a thermal insulation material coating on the outer wall, is filled with purified water to reflect radiation emissions to the PMTs. The PMTs are placed in the focal plane of the lens and arranged as a series of concentric ring matrixes supported by a stainless-steel space frame. The water lens array is designed to focus on the transient sources, such as GRBs, and AGNs (active galactic nuclei), in the northern sky to measure primary $\gamma$-rays in the energy range from 10 GeV to 100 TeV. Thanks to the properties of the water lens, it has an FOV of 0.84 sr ($30^{\circ}$ zenith angle), which is almost one order of magnitude larger than that of IACTs such as H.E.S.S., MAGIC, and CTA  \citep{bolmont2014camera, aleksic2016major, cta2011design}. It is more important that the effective area, angular resolution, and energy resolution at 300 GeV energy can reach 10000 m$^2$, 0.5$^0$, and 20$\%$, respectively, which is comparable to H.E.S.S. and MAGIC, except for the angular resolution. The water lens detection design can meet the observation requirements of our physical targets.
	
	The scintillator array is composed of 100 detectors and is designed to perform joint observations with a water lens.
	The installation and maintenance of a scintillator detector are particularly simple, and a scintillator detector can achieve a long-term stable performance as well as a good time resolution, which is widely used in extensive air shower (EAS) arrays \citep{lhaaso2010future}. The frame structure is made of stainless steel with an upside-down pyramidal shape, called a box of light guide (BLG). The interfaces of a BLG are light-tight, with a reflective material, Dupont Tyvek covering the inner surface with the purpose of increasing the efficiency of light collection significantly. A KD2000 plastic scintillator is adopted with dimensions of 1 m $\times$ 1 m $\times$ 2 cm and placed onto the top of the frame. A PMT with a diameter of 2 inches is chosen and installed on the bottom.  When charged particles induced by the primary CRs pass through the scintillator, photons are generated. Some of these photons can be refracted several times in BLG, and some of them finally can reach and be collected by the PMT. The main function of a scintillator detector is to measure the coincidence with the water lens and to confirm the performance of the water lens.

	\begin{figure*}[!htb]
		\centering
		\includegraphics[width=0.9\textwidth]{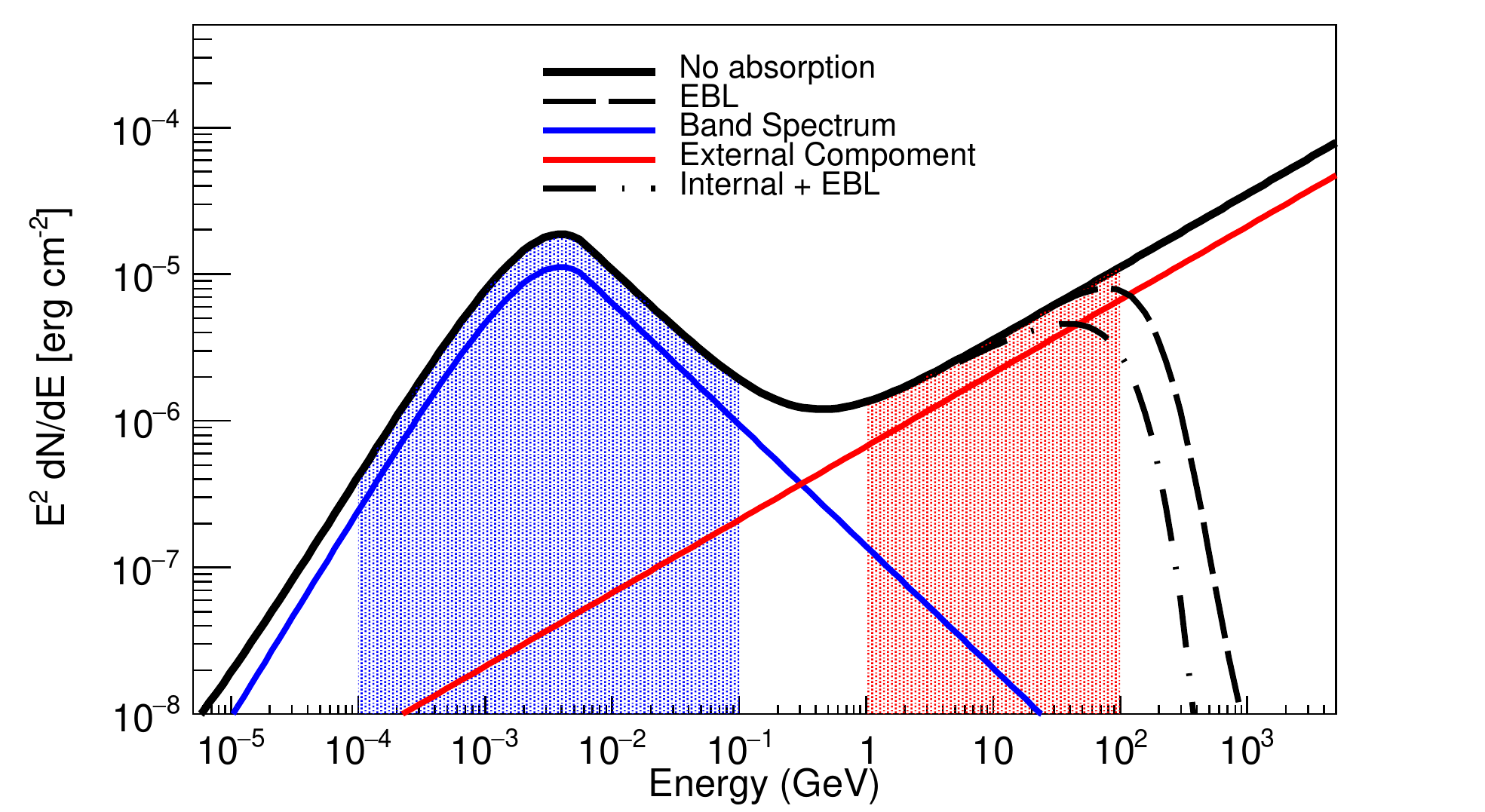}
		\caption{\label{fig:spectrum} The energy spectrum of the Band function with EBL and internal absorption is considered. The blue and red solid lines represent the spectra of the Band function and the extra component, respectively. The dashed and dotted lines represent the spectra when taking the EBL absorption of Ref.~\citep{finke2010modeling} and the internal + EBL absorption into account, respectively. The redshift is adopted as z = 0.903, and the cutoff energy is 50 GeV. The black lines are the sum of the Band function and the extra component. The values of $\beta_{ext}$ and $R_{ext}$ are -1.5 and 0.1, respectively. $R_{ext}$ represents the ratio of the fluences in the energy range of 100 MeV to 100 GeV to 100 keV to 100 MeV, which are the red-banded and blue-shaded regions, respectively. To distinguish the individual components of the emission from the total combined flux, both the Band spectrum and the extra component are scaled by 0.5. }
	\end{figure*}

	\section{\label{sec:SGRB}Modeling SGRB Samples}
	
	To estimate the detection rate of the prompt emissions from SGRBs for HADAR, we adopt a simulation method based on Monte Carlo. A set of SGRB samples is generated according to the phenomenological models of their intrinsic parameters, such as space density, luminosity, energy spectrum, temporal duration, and EBL attenuation. The expected detection rate is calculated using the spectra for each GRB and the HADAR sensitivity. 
	
	\subsection{Redshift Distribution}
	The redshift distribution describes the number of SGRBs per unit redshift bin $\mathrm{d}z$ per unit (observed) time $\mathrm{d}t$, which is written as \citep{zhang2018physics}:
	\begin{equation}
		\frac{\mathrm{d}N}{\mathrm{d}t\mathrm{d}z}=\frac{\dot{\rho}(z)}{1+z}\frac{\mathrm{d}V(z)}{\mathrm{d}z} \ ,
	\end{equation} 
	where 
	\begin{equation}
		\frac{\mathrm{d}V(z)}{\mathrm{d}z}=\frac{c}{H_0}\frac{4\pi D_L^2}{(1+z)^2[\Omega_m(1+z)^3+\Omega_{\Lambda}]^{1/2}} \ .
	\end{equation}
	In our work, we adopt $H_0 = 67.4\ \mathrm{km}\ \mathrm{s}^{-1}\ \mathrm{Mpc}$, $\Omega_m = 0.315$, and $\Omega_{\Lambda} = 0.685$ \citep{pdg2020}. In addition, $\dot{\rho}(z)$ is the event rate density at z (in units of $\ \mathrm{Gpc}^{-3} \ \text{yr}^{-1}$), which is \citep{howell2019joint,regimbau2009gravitational}:
	\begin{equation}
		\dot{\rho}(z)\propto(1+z)\int_{t_{min}}^{t_{max}}R_F(z_f)P(t_d)dt_d \ ,
	\end{equation}
	where $P(t_d)\propto1/t_d$,\ $t_{min}=20\ \mathrm{Myr}$ is the minimum delay time for a BNS system to evolve to merge, and $t_{max}$ is the age of the universe at the time of merging $t(z)$. $R_F(z_f)$ is the star formation history (SFH), which is adopted from Ref.~\citep{madau2014cosmic}.  We normalize the event rate density $\dot{\rho}(z)$ with $1540^{+3200}_{-1220}\ \mathrm{Gpc}^{-3} \ \text{yr}^{-1}$ \citep{abbott2017gravitational}, which corresponds to the local event rate density at $z = 0$. 
	
	The delay time $t_d$ between the  formation of the binary system $t_f(z_f)$ and the age of the universe at the time of merger $t(z)$ is given as 
	\begin{equation}
		t_d=\int_{z} ^{z_f}\frac{dz'}{H_0(1+z')[\Omega_m(1+z)^3+\Omega_{\Lambda}]^{1/2}} \ ,
	\end{equation}
	where $z_f$ and $z$ represent the redshifts at which the BNS systems form and merge, respectively.
	
	\subsection{Luminosity Function}
	We explore the expression of the SGRB's luminosity function in a broken power law, which is widely adopted in other works: 
	\begin{equation}
		\Phi(L) = A \left\{ \begin{matrix}
			(\frac{L}{L_c})^{\alpha}, & L \le L_c \\ 
			(\frac{L}{L_c})^{\beta}, & L > L_c
		\end{matrix} \right.
	\end{equation}
	$A$ is the normalization constant, $L_c$ is the break luminosity, and $\alpha$ and $\beta$ are the power-law indices. In this research,  we adopt  $\alpha = -1.95$, $\beta = -3$, $L_c = 2\times10^{52} \ \mathrm{erg}\  \mathrm{s}^{-1}, L_{min} =1\times10^{49} \ \mathrm{erg}\ \mathrm{s}^{-1}$\citep{wanderman2015rate} .

	\subsection{Prompt Emission Spectrum}
	The prompt emission spectrum of GRBs may include two spectral components: a non-thermal Band component (Band) and a non-thermal power law component extending to high energies (PL). In this work, we assume all of the pseudo-SGRBs have the Band + PL spectrum. 
	
	The Band function is usually used to fit the spectrum of a GRB when the detector's energy band is wide enough. The Band function is \citep{band1993batse, band2003comparison}: 
	\begin{eqnarray}
		N(E)=A_0 \begin{cases}
			\left( \frac{E}{E_0} \right)^{\alpha} \exp{\left(-E/E_p\right)}, & E \le E_c \\
			\left(\frac{E_c}{E_0}\right)^{\alpha-\beta}
			\exp{(\beta-\alpha)}\left(\frac{E}{E_0}\right)^{\beta}, & E > E_c
		\end{cases}
	\end{eqnarray}
	where $E_0=100 \ \mathrm{keV}$ and $E_c=(\alpha-\beta)E_p$. $A_0$ is the normalization constant in units of $\mathrm{photons} \ \mathrm{s}^{-1} \ \mathrm{cm}^{-1} \ \mathrm{keV}^{-1}$. $\alpha$ and $\beta$ are the low and high energy photon indices, respectively, and $E_p$ is the peak energy of the spectrum. In this work, $\alpha$ and $\beta$ are adopted from the observations of Fermi-GBM \citep{poolakkil2021fermi}. $E_p$ is determined by the $E_p-L_p$ relationships examined by Ref.~\citep{tsutsui2013possible}.
	
	Beyond the Band function, a high-energy and power law spectral component is necessary to fit the spectrum in some GRBs \citep{abdo2009limit,ackermann2010fermi,ackermann2011detection}. The power law component is the dominant contribution to the high-energy prompt emission beyond tens of GeV. In this work,  we introduce the extra component for GRB spectra. The luminosity ratio $R_{ext} = L_{ext} / L_{ave}$ and spectral index $\beta_{ext}$ are used to describe the extra component. $L_{ext}$ is the luminosity of the extra component and $L_{ave}$ is the luminosity of the Band component. We take $R_{ext} = 0.1$, $\beta_{ext} = 1.5$.

	\subsection{$\gamma \gamma $ Absorption}
	
	The photons with the highest energies within the emission region may be strongly attenuated by the low-energy photons through the $\gamma \gamma \rightarrow e^{\pm}$ interaction. The specific spectral breaks occur because of the two-photon pair production, which is related to different bulk Lorentz factors and the fireball radius \citep{chen2018constraints,asano2007prompt}. In this work, we introduce an exponential cutoff on the spectrum to describe the internal absorption when calculating the detection rate. For typical burst parameters, the internal shock spectrum cuts off above a threshold energy of $E_{\gamma,th}  \sim 10 - 100\ \text{GeV}$ and has no internal cutoff with a large enough bulk Lorentz factor \citep{razzaque2004gev}, so the cases of 30 GeV, 50 GeV, 100 GeV, and 1 TeV with and without an energy cutoff are considered. 
	
	High-energy photons from distant astrophysical sources are subject to attenuation because of the two-photon pair production with the EBL. This absorption of $\gamma$-rays can be described as $e^{-\tau(E,z)}$, where $\tau(E,z)$ is the optical depth for the $\gamma$-rays at energy E. The EBL attenuation introduced in the work of Ref.~\citep{finke2010modeling} is used by default. Figure~\ref{fig:spectrum} shows the energy spectrum of the SGRBs. The EBL and internal absorption are considered.

	\subsection{Temporal Properties}
	The duration of the GRBs is approximately estimated using $T_{90}$, which corresponds to the time in which $90\%$ of the counts arrive. $T_{90}$ is described as \citep{ghisellini2010gev,kakuwa2012prospects}:
	\begin{equation}\label{eqnT90}
		T_{90}=(1+z)\frac{E_{iso}}{L_{ave}} \ ,
	\end{equation}
	where isotropic energy $E_{iso}$ is calculated with \citep{kakuwa2012prospects}:
	\begin{equation}
		E_{iso} = 10^{51.42\pm0.15}\mathrm{erg} \left( \frac{E_p}{774.5 \mathrm{keV}} \right)^{1.58\pm0.28} \ .
	\end{equation}
	The average luminosity $L_{ave}$ is calculated by $L_{ave} = 0.31 L_p$. 
	
	\begin{figure*}[!htb]
		\centering
		\includegraphics[width=0.45\textwidth]{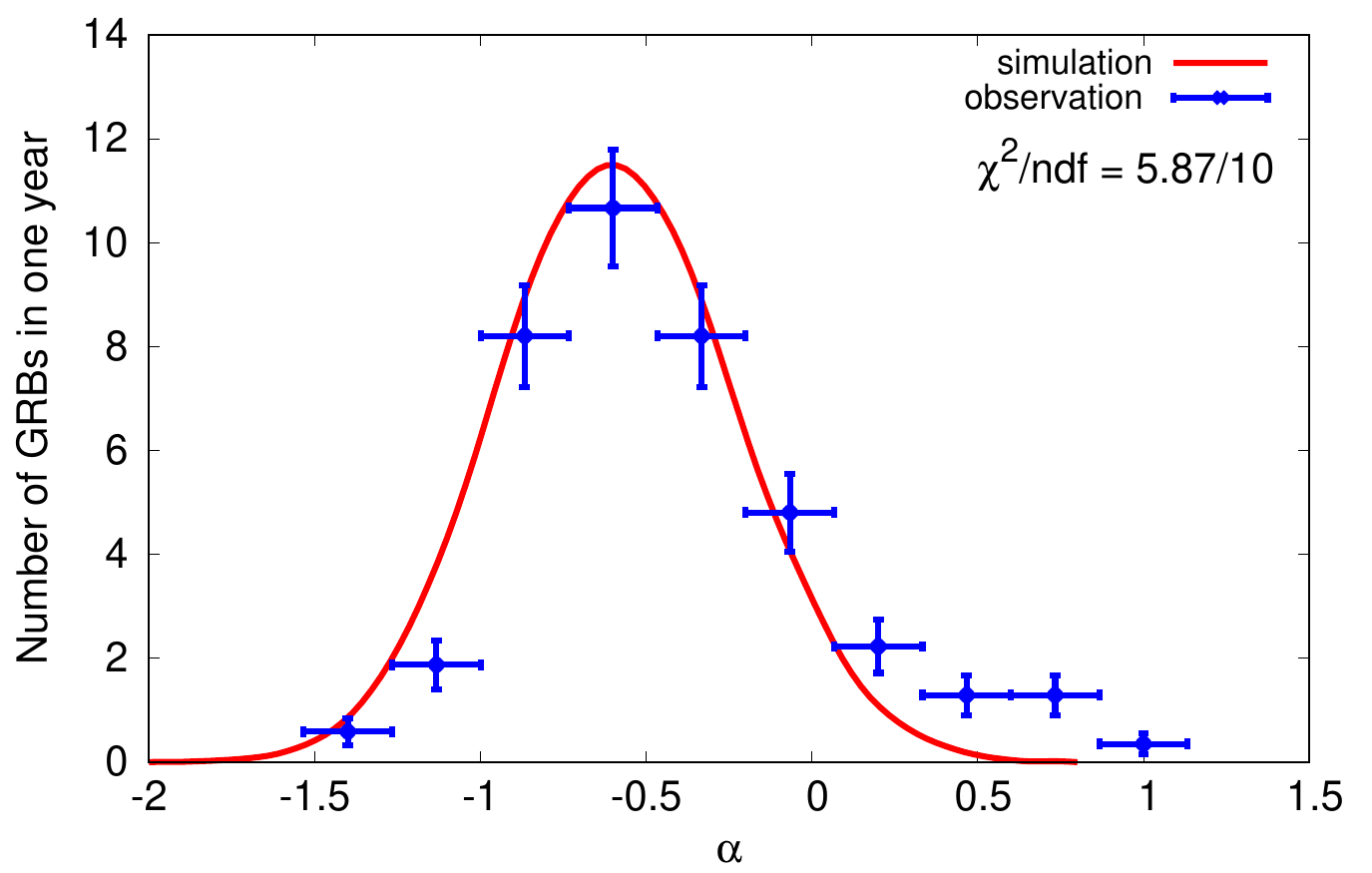}
		\includegraphics[width=0.45\textwidth]{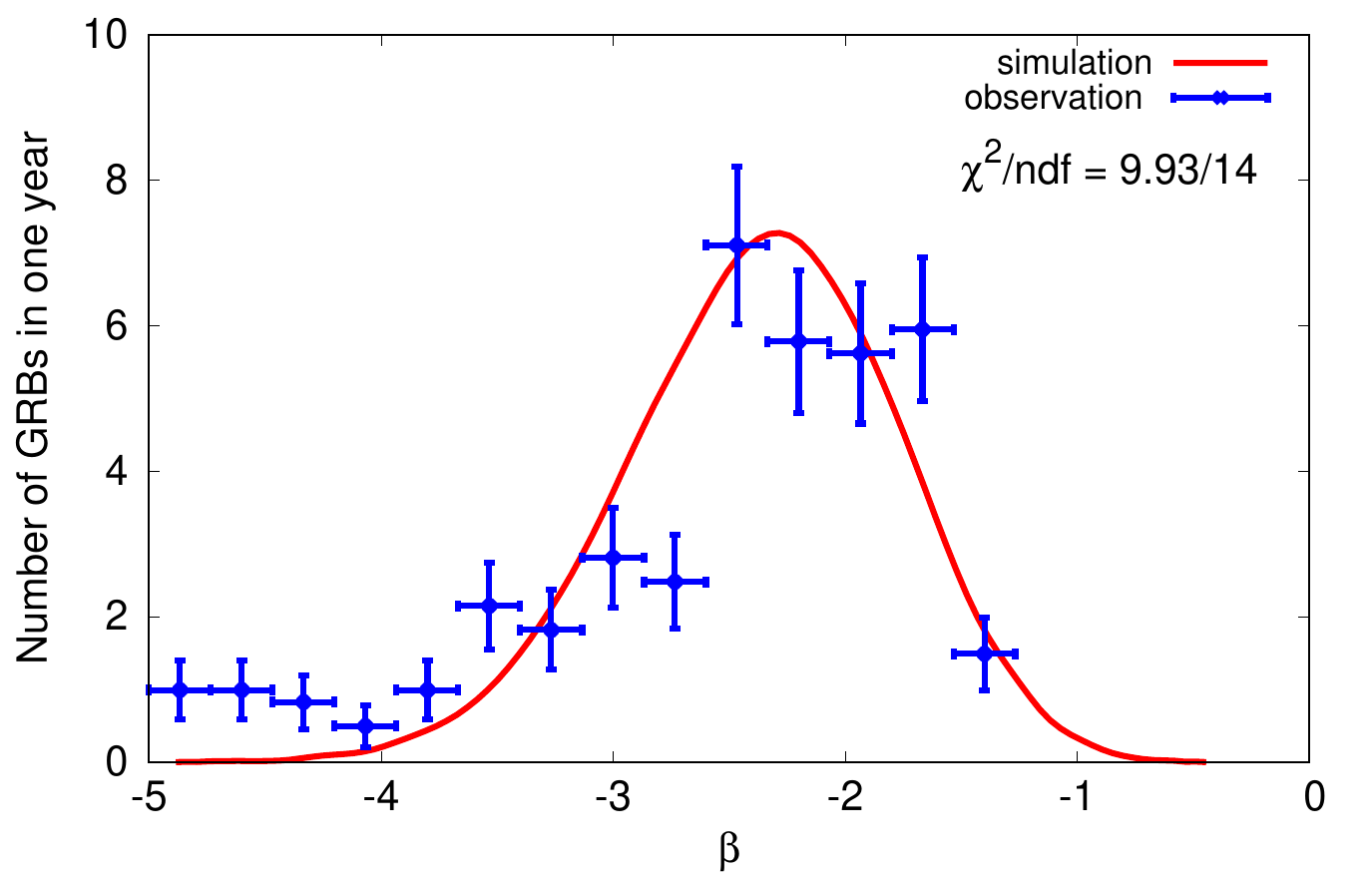}
		\includegraphics[width=0.45\textwidth]{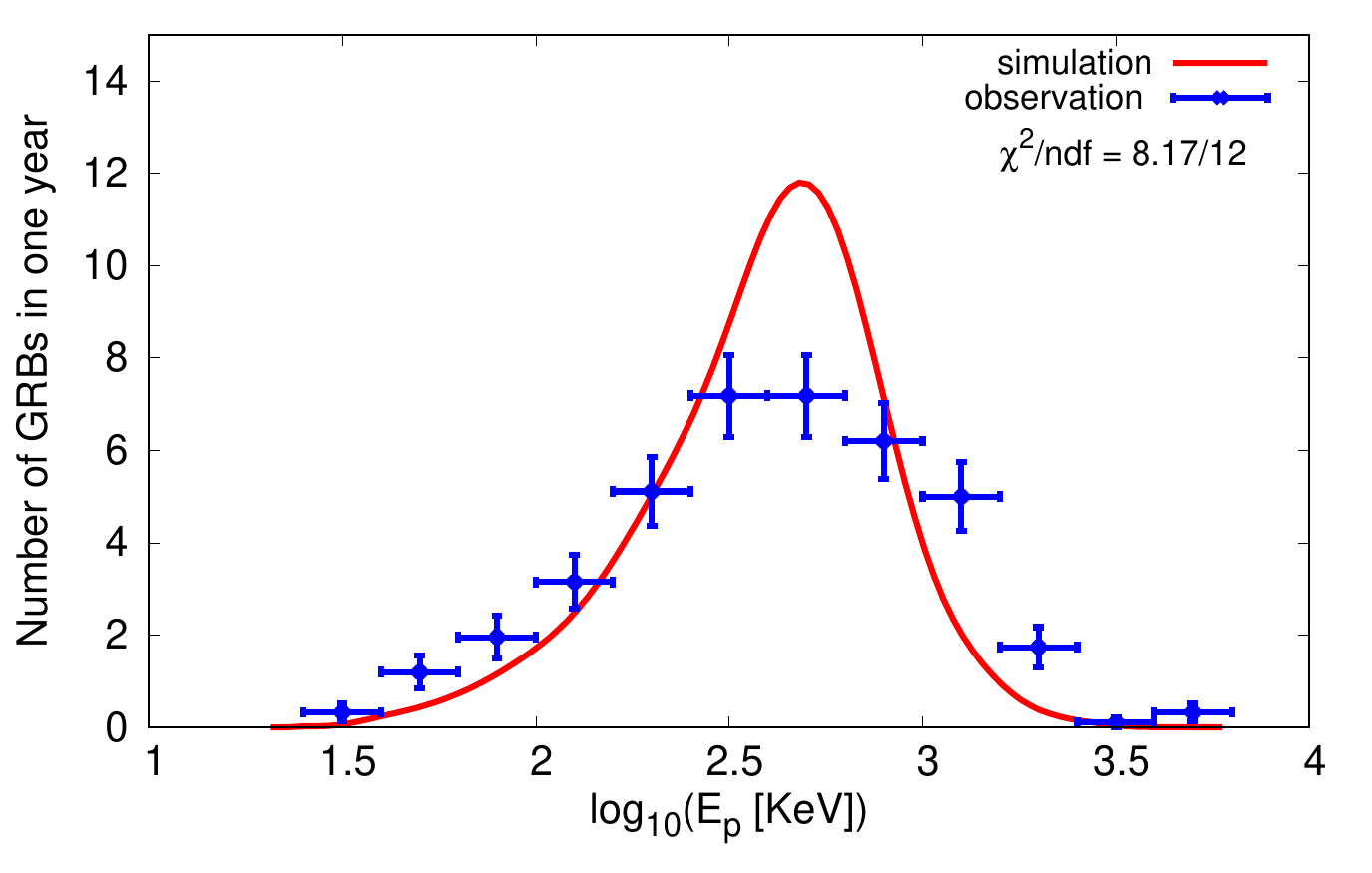}
		\includegraphics[width=0.45\textwidth]{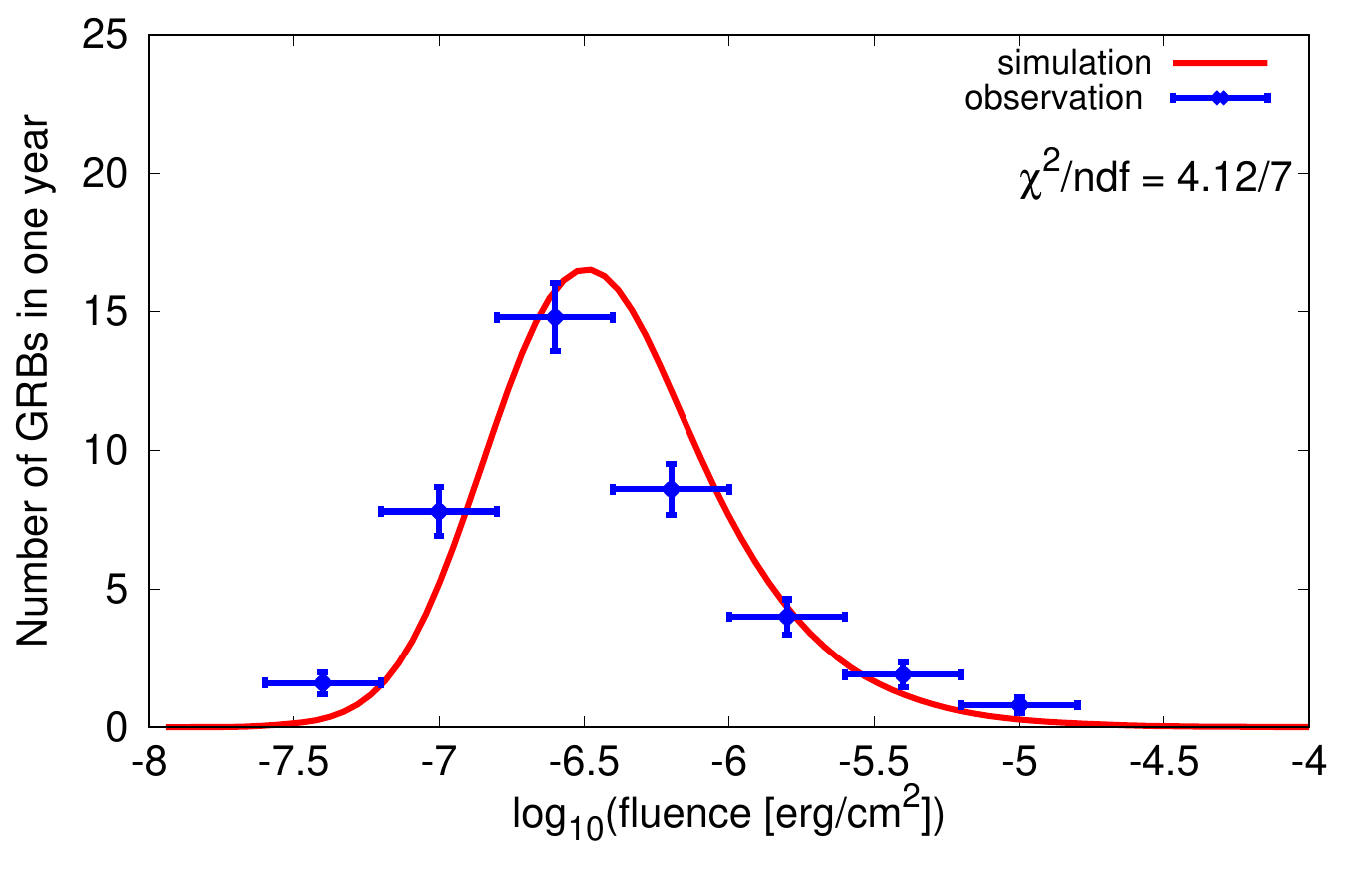}
		\includegraphics[width=0.45\textwidth]{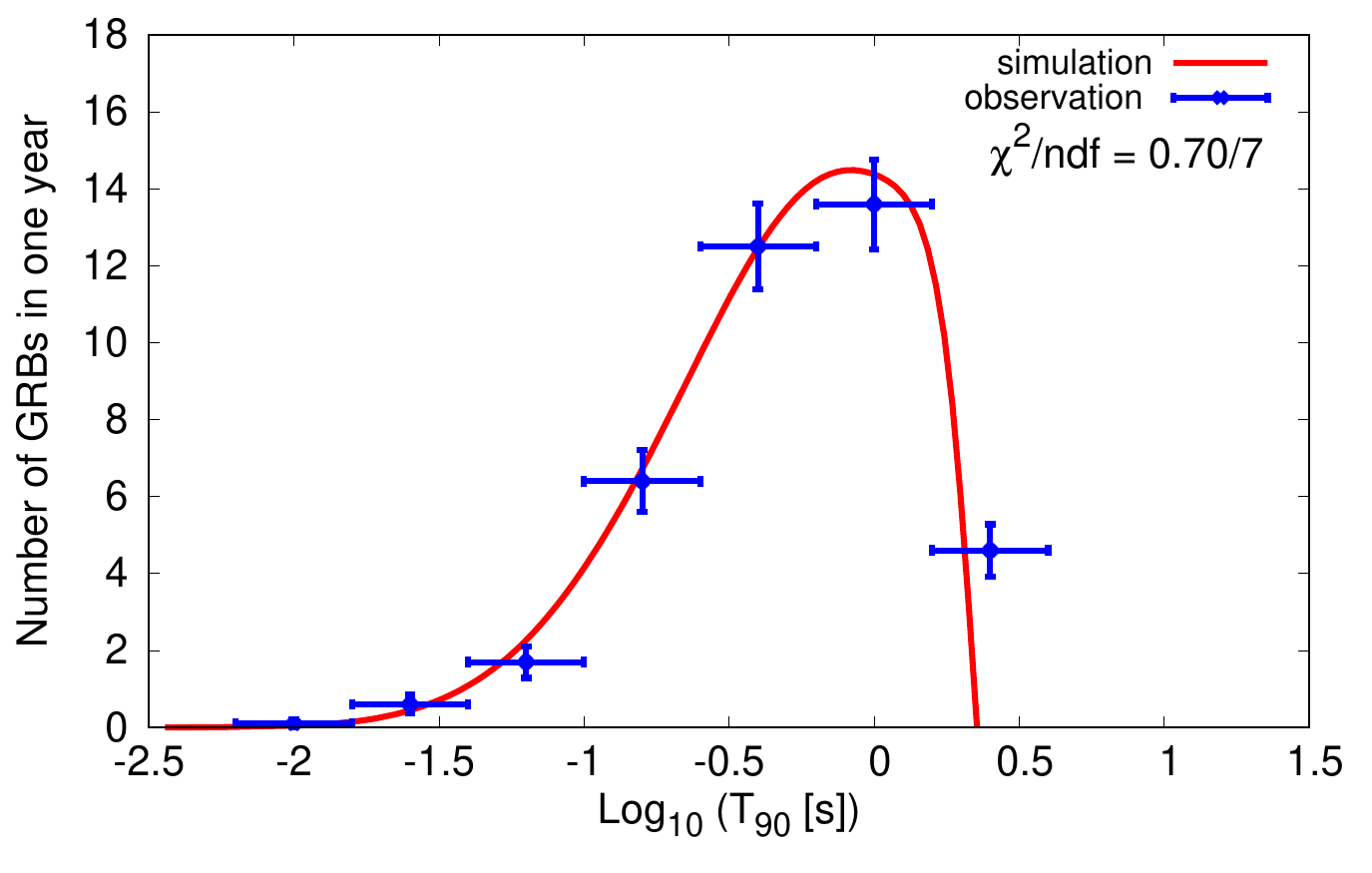}
		\includegraphics[width=0.45\textwidth]{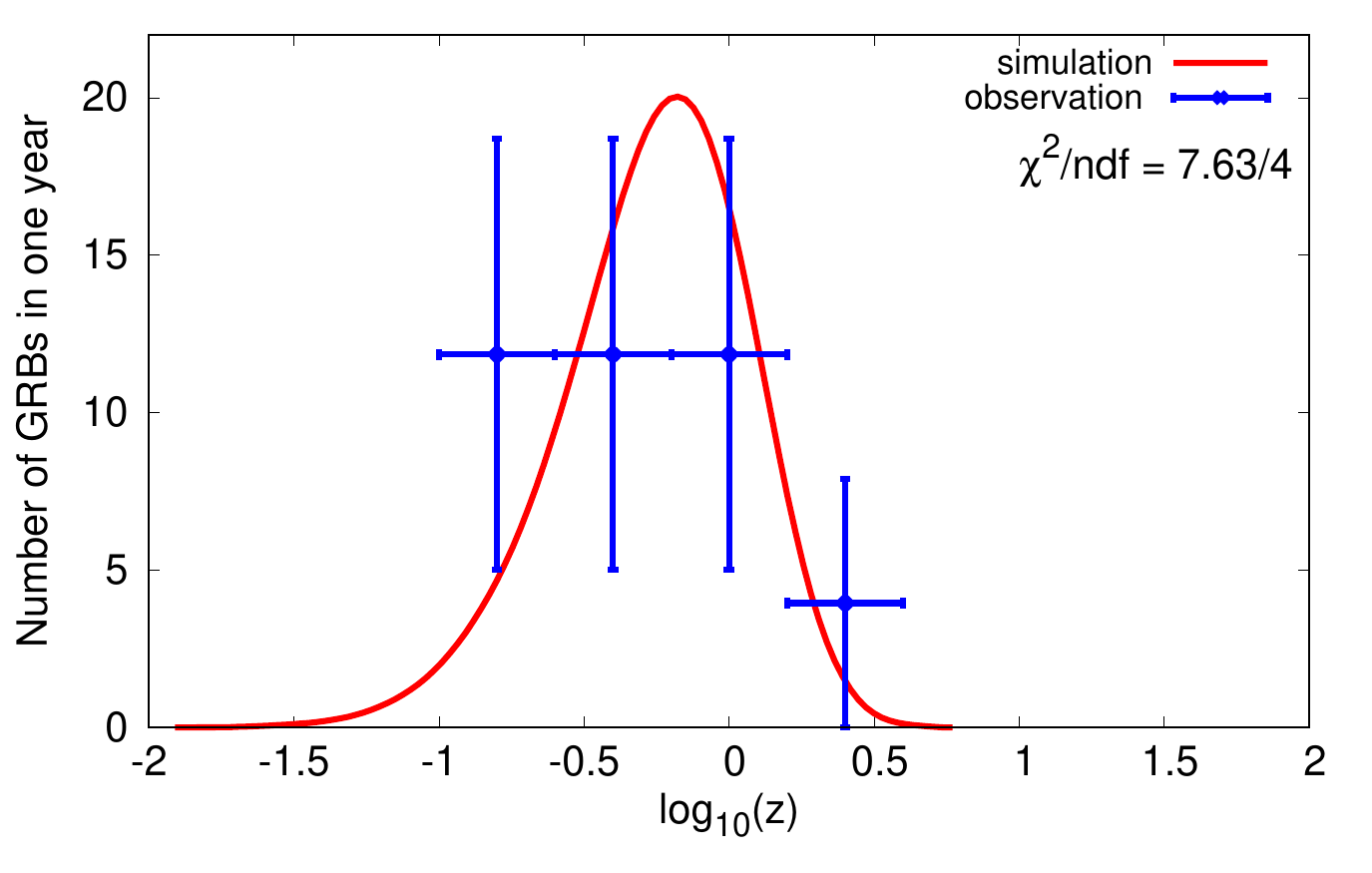}
		\caption{\label{fig:check} Model results of $\alpha$ (top left), $\beta$ (top right), $E_p$(middle left), fluence(middle right), $T_{90}$(bottom left), and redshift $z$ (bottom right).}
	\end{figure*}		
	
	\begin{figure*}[!htb]
		\centering
		\includegraphics[width=0.45\textwidth]{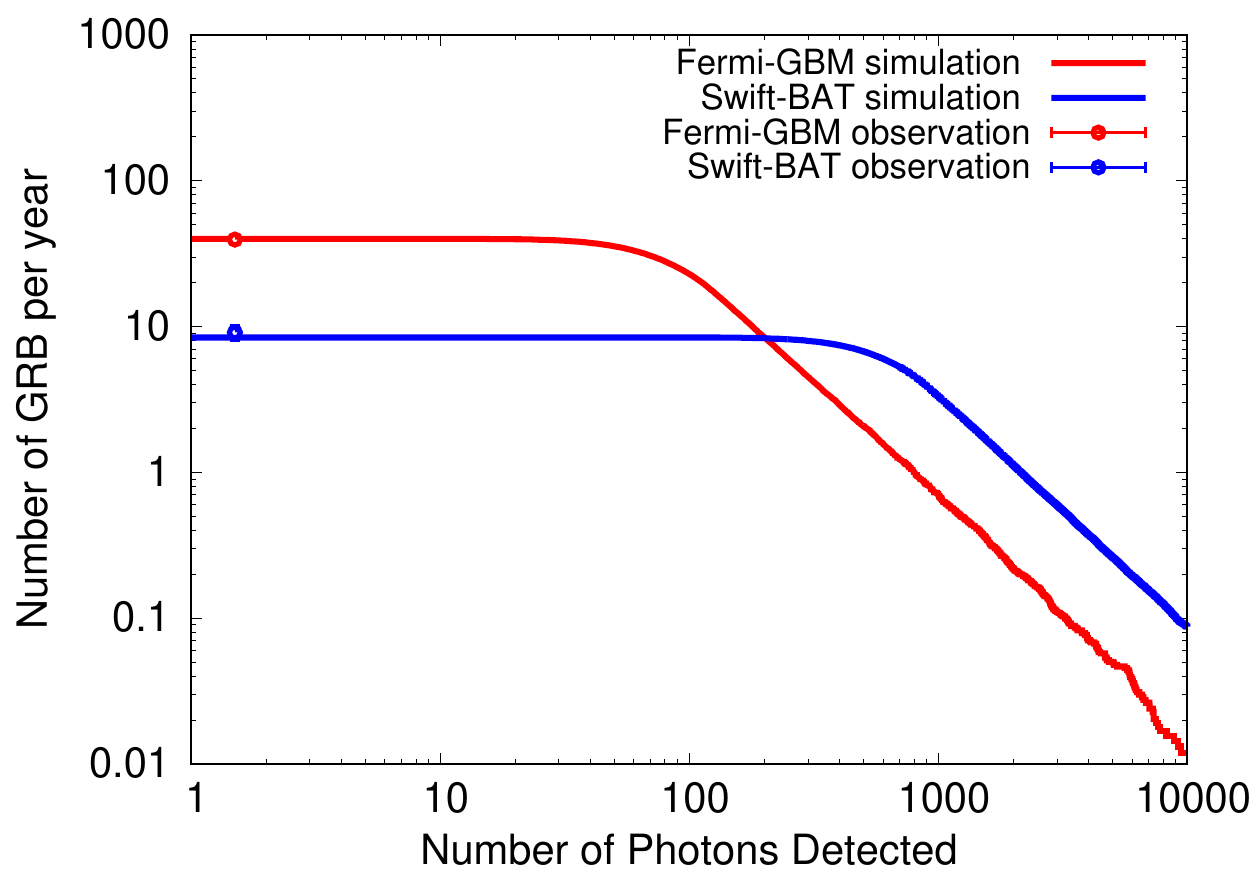}
		\includegraphics[width=0.45\textwidth]{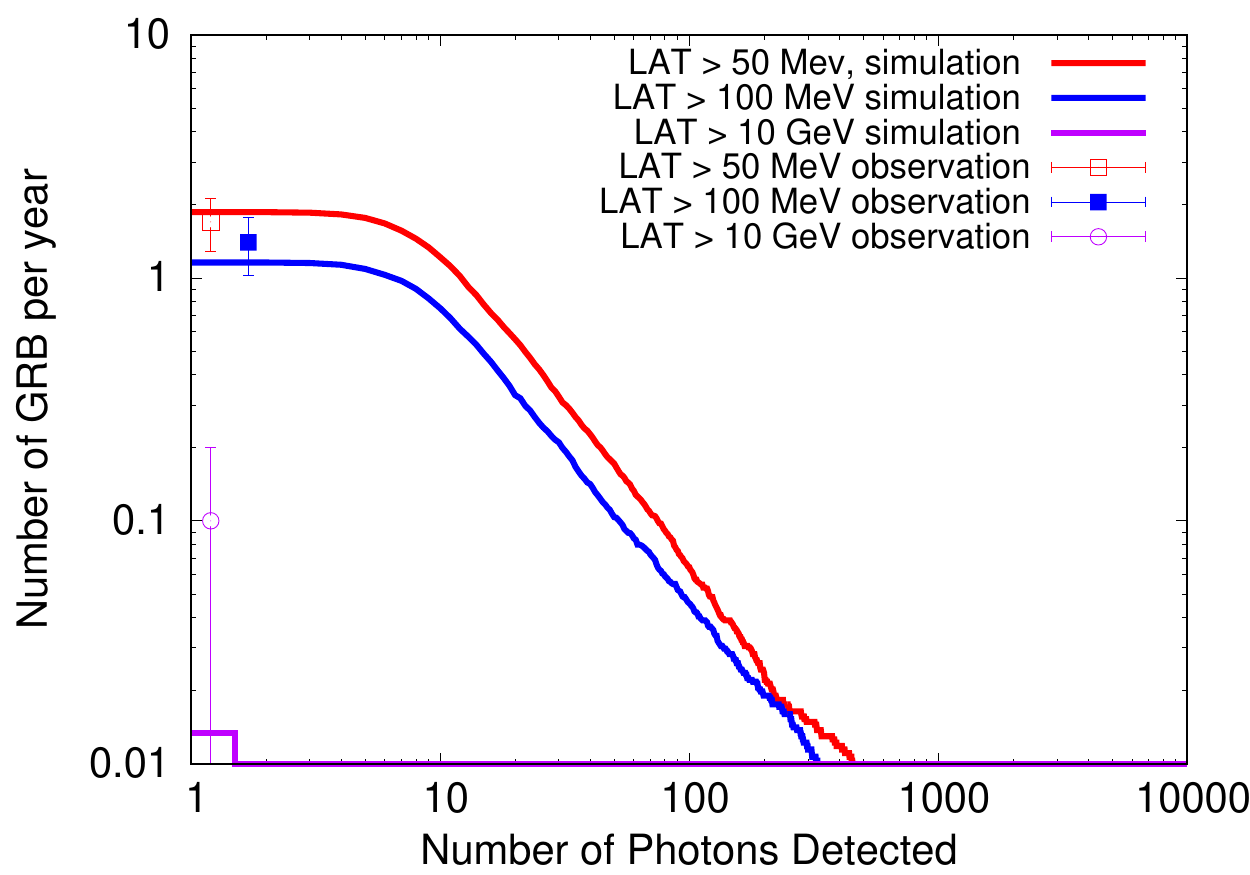}
		\caption{\label{fig:ratecheck} The simulated detection rates of SGRB with Fermi GBM, Swift BAT, and Fermi LAT. }
	\end{figure*}
	\subsection{Significance}
	Based on the preparations above, a group of GRBs can be constructed using Monte Carlo with the definite parameters $\left(  \alpha, \beta, E_p, L_p, T_{90}, z \right)$. Then the sample is ready for a quality check with Fermi-GBM observations. The significance for every GRB is calculated based on the expected signal and background number, $N_{signal}$ and $N_{bkg}$. The signal can be obtained as 
	\begin{equation}
		N_{signal}=\int_{E_{min}}^{E_{max}}N(E)T_{90}e^{-\tau(E,z)}A_{eff}^{\gamma}(E,\theta)\mathrm{d}E \ ,
	\end{equation}	
	where $T_{90}$ is described in Equation.~\eqref{eqnT90}. N(E) is the spectrum of the SGRB. $A_{eff}$ is the effective area of the detector. For HADAR, the effective area is adopted from Ref.~\citep{xin2021prospects}. The factor $e^{-\tau(E,z)}$ is from the EBL attenuation. The integral limit $(E_{min},E_{max})$ is the detector's energy band, which is $(10\ \text{GeV},10\ \text{TeV})$ for HADAR. 
	
	For space-bone detectors such as GBM, $N_{bkg}$ comes from cosmic $\gamma$ background rates in the corresponding energy bands.
	For HADAR, 
	$N_{bkg}$ results from the cascades of cosmic rays:
	\begin{equation}
		N_{bkg}=\int_{E_{min}}^{E_{max}}S_p(E)T_{90}A_{eff}^{p}(E,\theta)\Omega(E)\mathrm{d}E \ ,
	\end{equation}
	where $S_p(E)$ represents the spectrum of cosmic rays. $A_{eff}^p$ is the effective area of HADAR corresponding to cosmic rays, and $\Omega(E)$ is the angular resolution of HADAR.  These are given in Ref.~\citep{xin2021prospects}.
	
	A $5\sigma$ deviation is needed as a trigger, which is calculated as $N_{signal}/\sqrt{N_{bkg}} \ > 5$.

	\subsection{ Check by Observations}
	
	The observation of Fermi-GBM is used to check our samples, and the detection rates of Fermi-GBM, Swift-BAT, and Fermi-LAT are used to check our detection rate. In Fermi-GBM's observational results, 395 SGRBs are observed in ten years. To evaluate the fitting quality, $\chi^2_{tot}$ is calculated as $\sum \chi_i^{2}/\sum N_{i}^{bin}$, where $i$ denotes $\alpha, \beta, E_p, z, T_{90}$, and the fluence, and $N_i^{bin}$ is the corresponding bin number. The total combined $\chi^2_{tot}$ is 0.63.
	
	The $\alpha, \beta, E_p, z, T_{90}$, and fluence distributions are compared between our phenomenological models and real observations, as shown in Figure~\ref{fig:check}. From Figure~\ref{fig:check}, it can be seen that our model results are roughly consistent with the real observations of Fermi-GBM given the experimental uncertainties. 
	
	Figure~\ref{fig:ratecheck} shows the real detection rates of Fermi-GBM, Swift-BAT, and Fermi-LAT compared with our simulated detection rates.  We find that at a low energy or high energy, our simulated detection rates are roughly consistent with the real detection rates, given the experimental uncertainties.

	\begin{figure*}[!htb]
		\centering
		\includegraphics[width=0.45\textwidth]{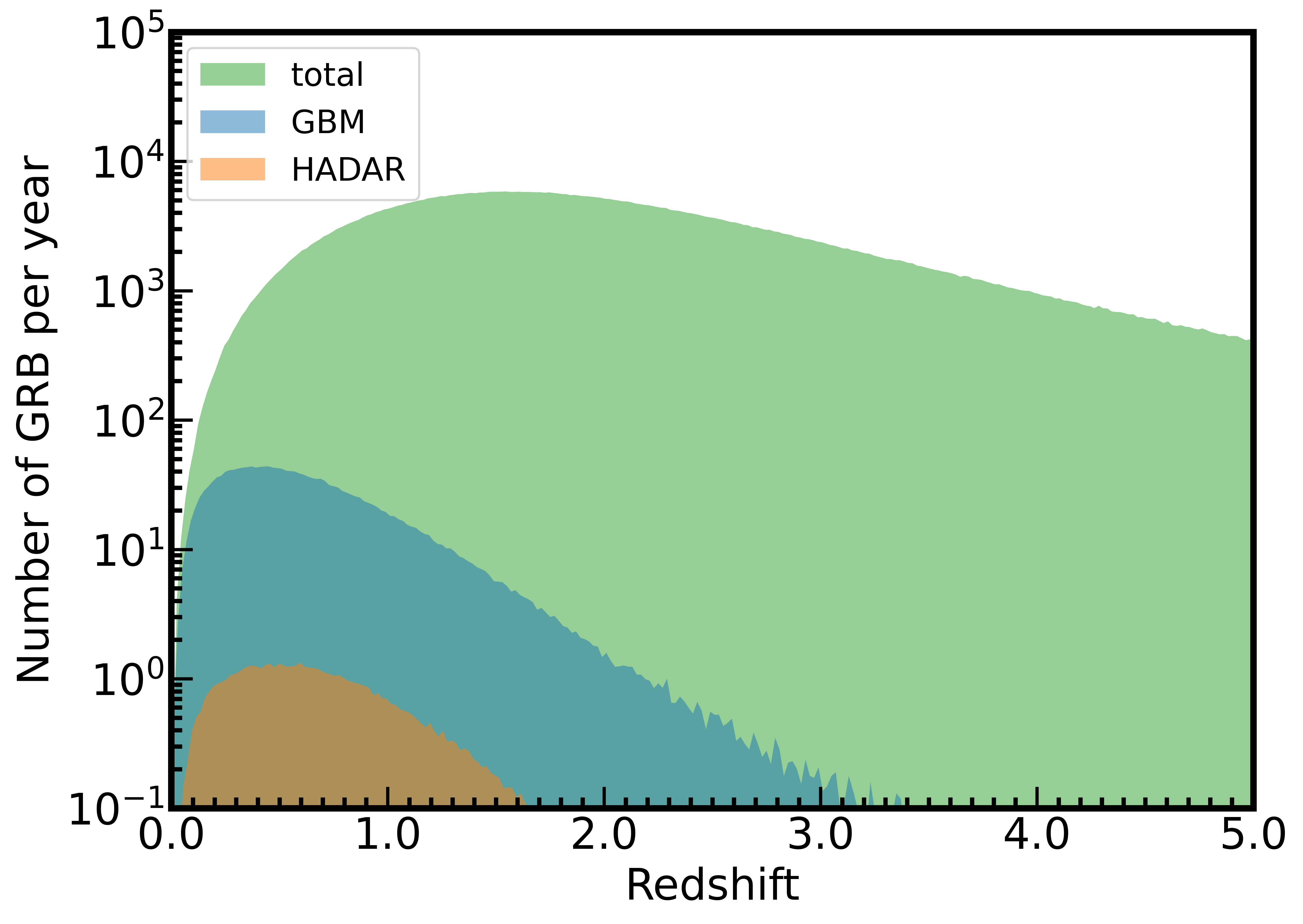}
		\includegraphics[width=0.45\textwidth]{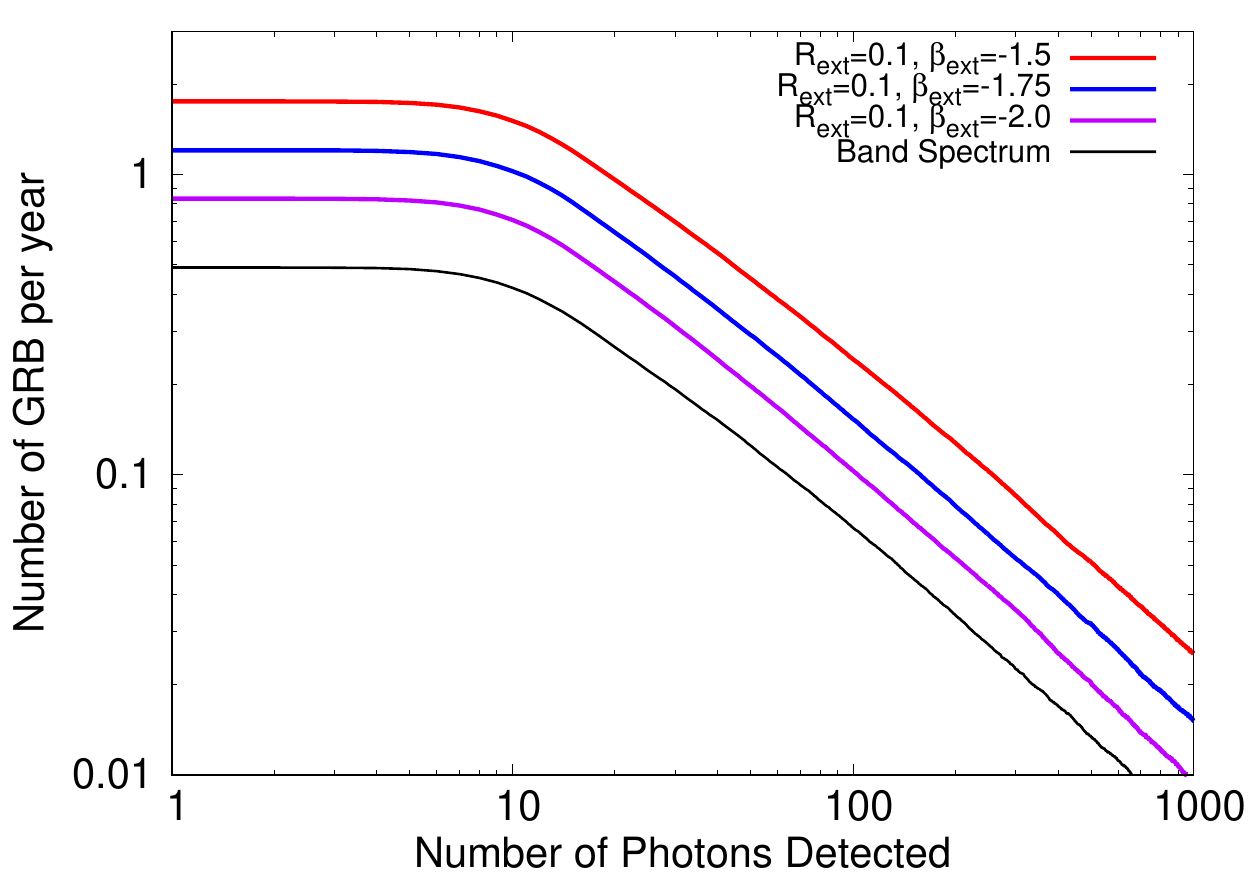}
		\caption{\label{fig:rate1} Left: The SGRB detection rate between the redshift z and dz of GBM and HADAR is shown by the blue and orange lines, while the green line shows the BNS merger rate between the redshift z and dz. Right: The SGRB dection rate of HADAR for different combinations or parameters for the extra component, in which the black line and the cyan line represent the Band spectrum, the yellow, blue, and red lines represent the combination of the extra + Band components. The model of Ref.~\citep{finke2010modeling} is considered, and no internal absorption is taken into account.}
	\end{figure*}

	\section{\label{sec:result}Results}
	
	The results of our work include two parts: the detected event rates of SGRBs, and a pseudo light curve for GRB090510 simulated with the HADAR experiment, which reflects HADAR's capacity to detect VHE SGRBs with enough sensitivity and a wide FOV and to observe photons extending to the VHE spectral component. 
	\begin{figure*}[!htb]
		\centering
		\includegraphics[width=0.45\textwidth]{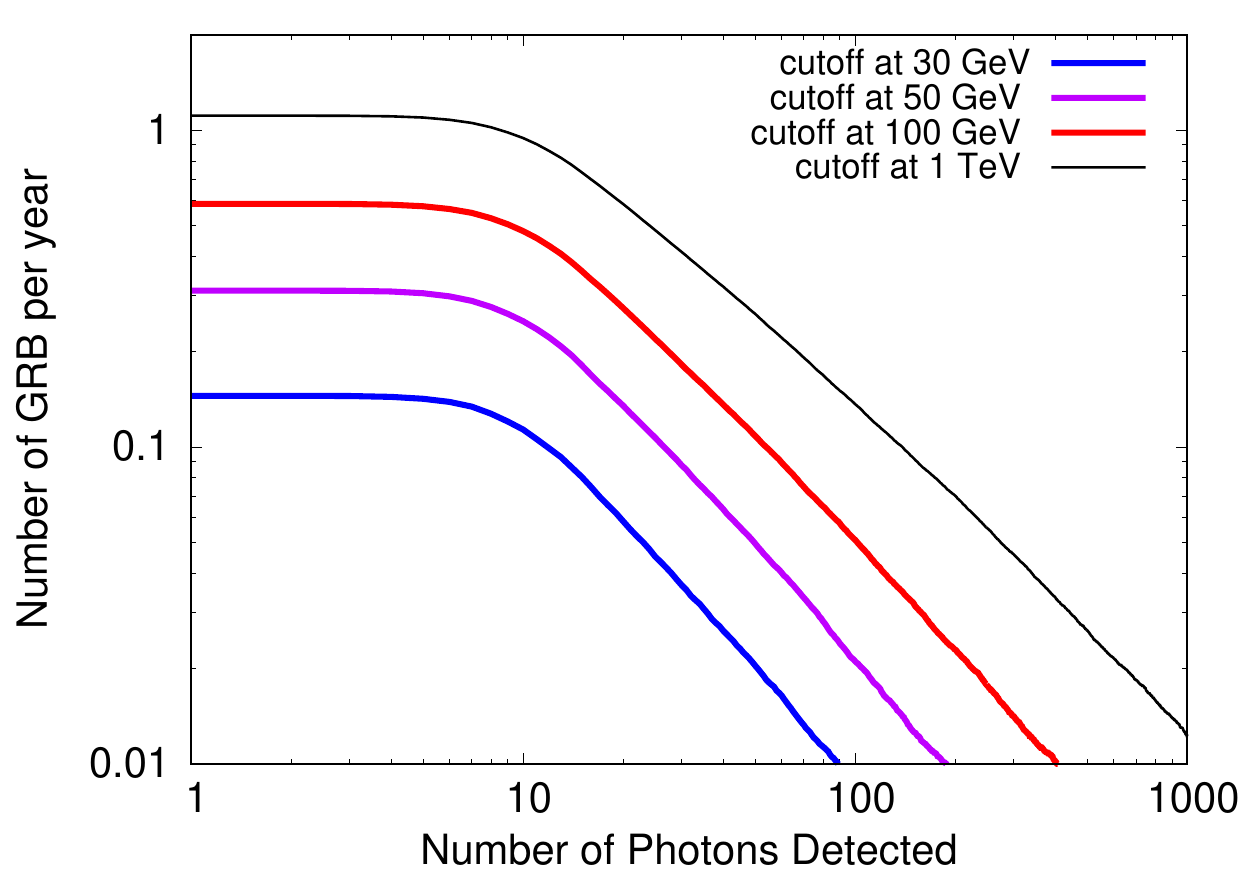}
		\includegraphics[width=0.45\textwidth]{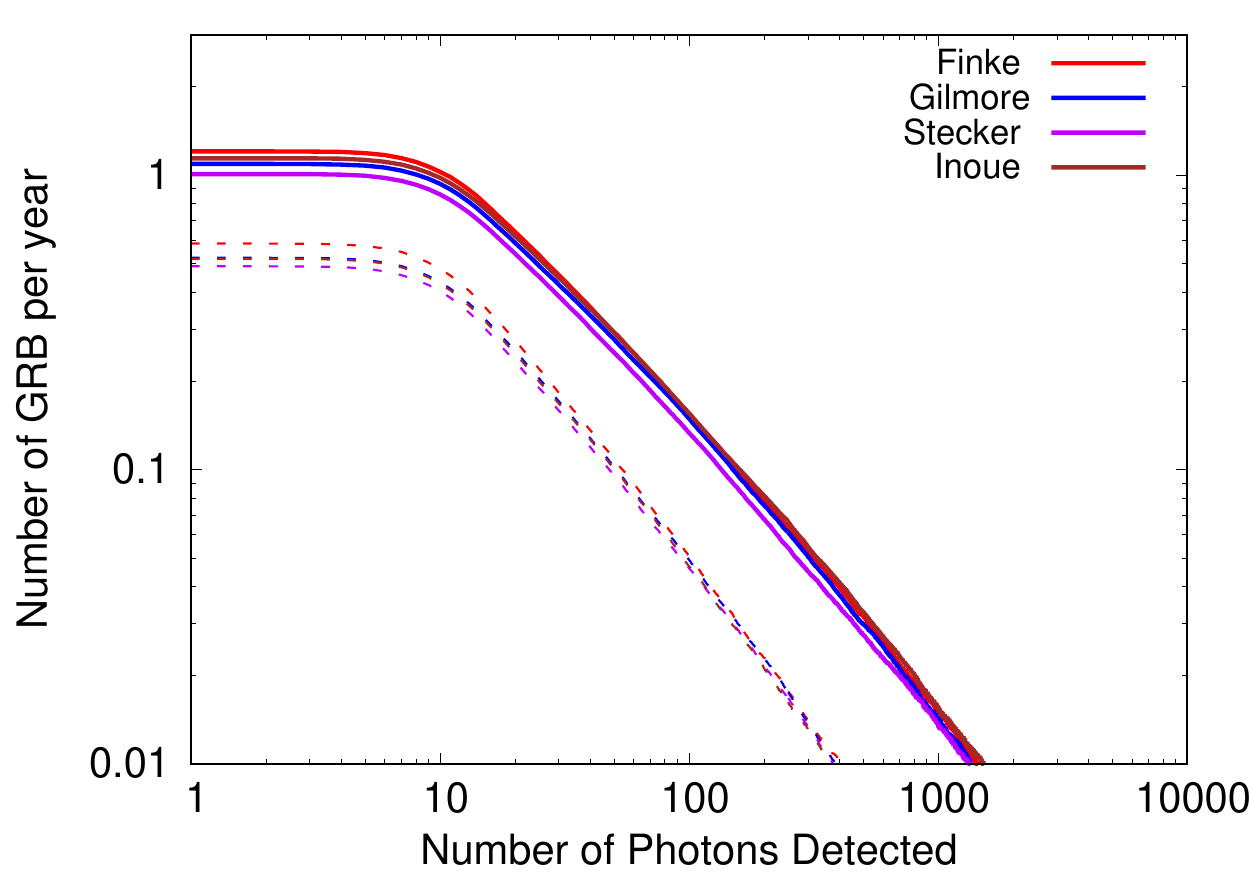}
		
		\caption{\label{fig:rate2}Top left: The SGRB detection rate with the extra + Band spectra considered for different energy breaks, in the case of $\beta_{ext}$ = -1.5, $R_{ext}=0.1$, and the EBL model \citep{finke2010modeling}. Top right: The SGRB detection rate with the extra + Band spectra considered for different EBL models, $\beta_{ext}$ = -1.5, $R_{ext}=0.1$. The solid lines are cases without internal absorption and the dashed line are the results when the energy spectrum is considered to cut off at 100 GeV. }
	\end{figure*}
	
		\begin{figure*}[!htb]
		\centering
		\includegraphics[width=0.9\textwidth]{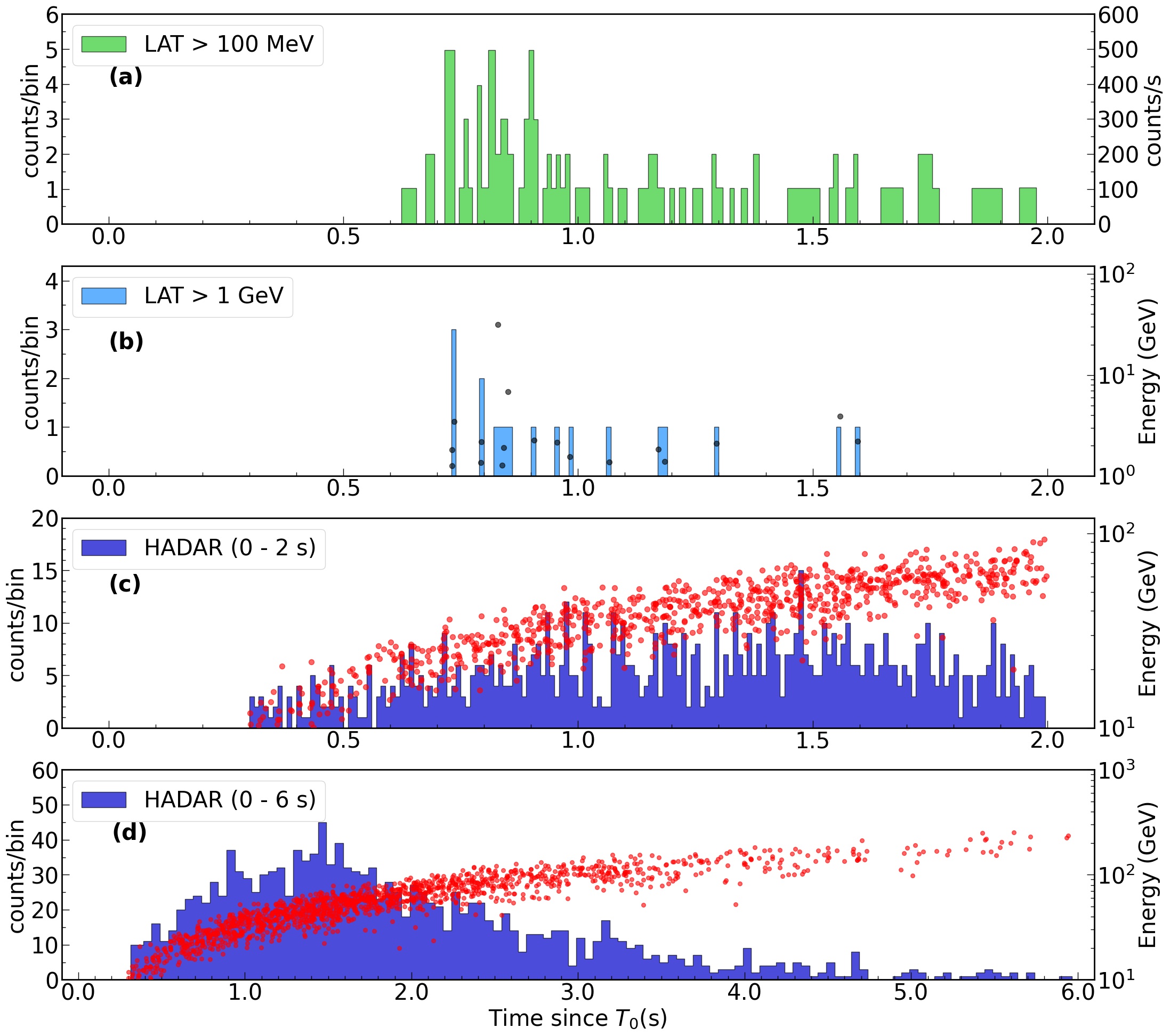}
		\caption{\label{fig:LC}Bottom: Light curves of GRB090510 at different energies. \texttt{a}, LAT light curve with E $<$100 MeV. The per-second count rate is displayed on the right side for convenience. \texttt{b}, LAT light curve with E $<$1 GeV. \texttt{c}, a simulated light curve of 090510-like GRB detected by HADAR. \texttt{b} and \texttt{c} also overlay the energy versus arrival time for each photon, with the energy scale displayed on the right side. \texttt{d}, the full light curve described in \texttt{c}. Photons are represented as gray and red dots in \texttt{b} and \texttt{c} separately.}
	\end{figure*}
	\subsection{\label{sec:eventrate} Expected Detection Rates}
	The left panel of Figure~\ref{fig:rate1} presents the expected detection rate between $z$ and $z+\mathrm{d}z$ for GBM and HADAR compared with the BNS merger rate. The right panel of
	Figure~\ref{fig:rate1} presents the model calculation of the SGRB detection rate for HADAR, which is the number of detected SGRBs as a function of the number of detected photons. The lines represent the extra spectra for the spectrum, $R_{ext} = 0.1$ and $\beta_{ext}=-2.0,-1.75,-1.5$. It is evident that the detection rate for HADAR is 0.49 per year for the Band spectrum. When the extra component with $R_{ext} = 0.1$ is considered, the detection rates for HADAR increase to 0.83, 1.2, and 1.8 for $\beta_{ext}=-2.0,-1.75,-1.5$.

	Considering two-photon pair production within the emission region and during the propagation, the results are presented in Figure~\ref{fig:rate2}. It can be seen that the internal absorption prevents HADAR from observing the VHE photons to some extent. If the energy cutoff due to the two-photon pair production within the emission region is higher than 30 GeV, HADAR can detect at least 0.15 SGRB per year with the EBL model of Ref.~\citep{finke2010modeling}. The result in the right-hand panel of Figure~\ref{fig:rate2} shows that the different EBL models have a relatively small impact on the detection rate compared with that of the internal absorption. Adopting different EBL absorption models from Refs.~\citep{finke2010modeling,gilmore2012semi,stecker2012determination,inoue2013extragalactic}, the expected detection rate of HADAR is approximately 0.49--1.1 SGRBs per year with or without a 100 GeV energy cutoff.

	\subsection{\label{sec:lightcurve}Pseudo Light Curve of GRB090510}
	Based on our HADAR model and the parameters of GRB090510 extracted from observational results \citep{ackermann2010fermi}, including the redshift and spectrum, taking into account the model of the LIV effects for this GRB from Ref. ~\citep{abdo2009limit}, a pseudo light curve of HADAR observation is generated, as shown in the lower panels (\texttt{c}) and  (\texttt{d}) of Figure~\ref{fig:LC}, as well as each photon with the observed time and energy. 
	
	To display this light curve, we consider the photon energy distribution to follow the spectrum of 090510-like GRB combined with EBL attenuation. Then, these photons are detected by HADAR with effective areas and a $20\%$ energy resolution. An assumption is made that all photons from the source burst at exactly the same time, which means that the duration for the emission at the source is $\Delta t = 0$. For the LIV effects, a positive time delay $\Delta t = (\Delta E/E_{QG})D/c$ is included in the simulated arrival time of photons, where $E_{QG}=M_{QG}c^2$ is the Planck energy, D is the co-moving distance from source to detector, and $\Delta E = E_{\gamma, high} - E_{\gamma,low}$. The energy dispersion of the time delay $\Delta t/\Delta E = +0.03\ \text{s}/\text{GeV}$ is adopted \citep{abdo2009limit}. From Figure~\ref{fig:LC}, it is clear that HADAR works better in higher-energy photon detection than LAT, which is helpful for providing more constraints on the parameters of LIV effects.

	\section{\label{sec:conclusion}Conclusions}
	The observations of TeV $\gamma$-ray emissions from SGRB are very important for studying the central engine, radiation mechanism, and the new physics of LIV particularly. Limited by the effective area and narrow FOV of space-borne and ground-based applications with IACT technology experiments, there are still no results. The new experiments are required to have a wide FOV and high sensitivity with IACT technology. The HADAR experiment can fulfill these two characteristics simultaneously. 
	
	In this work, we produce one SGRB sample and then check the sample with Fermi-GRM, Fermi-LAT, and SWIFT observations. Based on this sample, the annual detection rate of SGRB is calculated for HADAR experiments. For the present status, the HADAR can detect 0.5 SGRB in one year. The annual rate can slightly change with different EBL absorption models and intrinsic spectral break-off. Furthermore, we assume that GRB090510A-like events appear in the FOV of HADAR. HADAR can detect about 2000 photons with the internal absorption considered. Simultaneously, the evolution effect of the light curve increases with the energy, which can be clearly observed when LIV happens in the Planck energy. Based on this work, we believe that the HADAR experiment has a good performance in the observation of SGRBs. We hope that the HADAR experiment will be successfully completed and that it can observe SGRBs that are consistent with our expectations in the future.

	\acknowledgments
	
	This work is supported by the National Natural Science Foundation of China (12263004, 12263005, and 12275279).
	

	\bibliographystyle{aasjournal}
	\bibliography{ref}
	
\end{document}